\documentclass[aps,prd,nofootinbib]{revtex4}
\usepackage[colorlinks=true, pdfstartview=FitV, linkcolor=blue, citecolor=red, urlcolor=magenta]{hyperref}
\usepackage{graphicx}
\usepackage{latexsym}
\usepackage{amsmath}
\usepackage{amsfonts}
\usepackage{amssymb}
\usepackage{verbatim}
\usepackage{extarrows}

\newcommand{\be}{\begin{equation}}
\newcommand{\ee}{\end{equation}}
\newcommand{\bea}{\begin{eqnarray}}
\newcommand{\eea}{\end{eqnarray}}

\newcommand{\ben}{\begin{eqnarray}}
\newcommand{\een}{\end{eqnarray}}

\begin{document}

\title{Thermal Casimir effect in the Einstein Universe with a spherical boundary}
%Thermal corrections to the scalar vacuum energy density in spacetimes with nontrivial topology
%%%%%%%%%%%%%%%%%%%%%%%%%%%%%%%%%%%%%%%%%%%%%%%%%%%%%%%%%%

\author{$^{1}$ H. F. S. Mota}
\email{hmota@fisica.ufpb.br}

\author{$^{2}$C. R. Muniz}
\email{celio.muniz@uece.br}

\author{$^{1}$ V. B. Bezerra}
\email{valdir@fisica.ufpb.br}

%\affiliation{$^{1}$Departamento de F\'{\i}sica, Universidade Federal de Campina Grande,\\
%%Caixa Postal 10071, 58429-900, Campina Grande, Para\'{\i}ba, Brazil.}
\affiliation{$^{1}$Departamento de F\' isica, Universidade Federal da Para\' iba,\\  Caixa Postal 5008, Jo\~ ao Pessoa, Para\' iba, Brazil.}
\affiliation{$^{2}$ Universidade Estadual do Cear\'a, Faculdade de Educa\c{c}\~ao, Ci\^encias e Letras de Iguatu, Av. D\'{a}rio Rabelo, s/n,
63500-000, Iguatu, CE, Brazil.}
%%%%%%%%%%%%%%%%%%%%%%%%%%%%%%%%%%%%%%%%%%%%%%%%%%%%%%%%%%

%\preprint{}

\begin{abstract}
In the present paper we investigate thermal fluctuation corrections to the vacuum energy at zero temperature of a conformally coupled massless scalar field whose modes propagate in the Einstein universe with a spherical boundary, characterized by both Dirichlet and Neumann boundary conditions. Thus, we generalize the results found in literature in this scenario, which has considered only the vacuum energy at zero temperature. To do this, we use the generalized zeta function method plus Abel-Plana formula and calculate the renormalized Casimir free energy as well as other thermodynamics quantities, namely, internal energy and entropy. For each one of them we also investigate the limits of high and low temperatures. At high temperatures we found that the renormalized Casimir free energy presents classical contributions, along with a logarithmic term. Also in this limit, the internal energy presents a classical contribution and the entropy a logarithmic term in addition to a classical contribution as well. Conversely, at low temperatures, it is shown that both the renormalized Casimir free energy and internal energy are dominated by the vacuum energy at zero temperature. It is also shown that the entropy obeys the third law of thermodynamics. 
\end{abstract}
%\pacs{11.15.-q, 11.10.Kk}
 \maketitle

%\vspace{1cm}

%%%%%%%%%%%%%%%%%%%%%%%%
\section{Introduction}
\label{intro}
%%%%%%%%%%%%%%%%%%%%%%%%
%
The Einstein universe is a particular solution of the general relativity equations, which originally describes a spatially closed universe with constant positive curvature and uniform distribution of matter on the largest scales. It is in fact the Friedman-Robertson-Walker (FRW) spacetime that characterizes an universe under the assumption of the cosmological principle, but with a constant scale factor \cite{Ford:1975su, Ford:1976fn, Bezerra:2011zz, Bezerra:2011nc, Ozcan:2001cr, Bayin:1993yq, Kennedy:1980kc}. Hence, the Einstein universe is a model describing a closed static universe.

Such a cosmological model is certainly idealized, considering recent observations related to the current universe geometry, which appears to be flat in its overall scale \cite{Planck:2018vyg}. However, more realistic cosmological scenarios where closed static universes play a central role are described by emergent models which consider the Einstein universe as the suited geometry to characterize the stage preceding the inflationary epoch. During this age, close to the Planck era, the non-negligible vacuum fluctuations associated with quantum relativistic fields may provide robust contributions to determine parameters for those models \cite{Ellis:2002we, Ellis:2003qz}. Furthermore, modified theories that point towards the quantum gravity has also been investigated taking into consideration the Einstein universe geometry \cite{Boehmer:2009fey, Boehmer:2007tr, Li:2013xea, Bezerra:2017zqq, Shabani:2016dhj}.

The Casimir effect is a phenomenon associated with vacuum fluctuations of quantum-relativistic fields, and was originally thought of as caused by modifications, in Minkowski spacetime, of the zero-point oscillations of the electromagnetic field due to the presence of two ideal parallel conducting plates, at zero temperature \cite{Casimir1948dh}. The present status of the phenomenon is that it can also occur due to fields such as the scalar, spinor, gravitational, as well as unusual fields such as the   `Elko' one - a 1/2 spin neutral fermion field with mass dimension one in (3 + 1)-dimensional spacetime \cite{Maluf:2019ujv}. One can also take into account in the phenomenon the thermal influences and nontrivial topologies associated with different geometries of the spacetime (for reviews, see \cite{Milton:2001yy, bordag2009advances}). As an iconic example, we cite wormholes \cite{Khabibullin:2005ad, Sorge:2019kuh, Garattini:2019ivd, Santos:2020taq, Santos:2021jjs, Alencar:2021ejd}, in which exotic negative energies like the Casimir one would be even more necessary in order to stabilize them.

The Casimir effect arising as a consequence of the nontrivial topology of the spacetime has been considered, for instance, in Refs. \cite{Dowker:1978vy, DeWitt:1979dd, Lima:2006rr}. In the static Einstein universe (as well as in the FRW one), the effect has been investigated in a variety of works \cite{Ford:1975su, Ford:1976fn, Bezerra:2011zz, Bezerra:2011nc, Ozcan:2001cr, Bayin:1993yq, Kennedy:1980kc, Zhuk:1996xc, Bezerra:2014pza, Mota:2015ppk, Bezerra:2016qof, Bezerra:2021qnw, Herdeiro:2007eb}, including thermal correction analysis. The Casimir effect in the Einstein universe was also investigated in the presence of a spherical boundary \cite{Bayin:1993yq, Kennedy:1980kc}, as well as in a D-dimensional spacetime \cite{Herdeiro:2007eb, Ozcan:2001cr}. In general, the role of the quantum vacuum energy in the primordial universe is highlighted in such scenarios. Therefore, as the early stages of the universe are known to have been extremely hot, an investigation of thermal corrections to the vacuum energy at zero temperature in the Einstein universe becomes of great interest. The Casimir effect is also important to be explored in curved spacetimes, like the one intended here, from the fundamental point of view, with the aim to investigate the role played by the spacetime nontrivial topology.
In this work, we investigate thermal fluctuation corrections to the vacuum energy at zero temperature of a conformally coupled massless scalar field whose modes propagate in the Einstein universe with a spherical boundary.  The latter is characterized by both Dirichlet and Neumann boundary conditions. We, thus, generalized the results found in Refs. \cite{Bayin:1993yq, Kennedy:1980kc} where the authors considered the vacuum energy at zero temperature by adopting the cutoff method. We show that there exists remarkable differences between thermal Casimir quantities obtained in these two cases, unlike the scenario of zero temperature analysed in Refs. \cite{Bayin:1993yq, Kennedy:1980kc}, where the vacuum energy is the same for both boundary conditions mentioned.

The structure of the paper is as follows: In Section II, we present the Einstein universe spacetime with the spherical boundary and the zeta function method to calculate the temperature corrections. In Section III, we obtain finite temperature contributions that must be subtracted in order to define the renormalized Casimir free energy. In Section IV, we calculate the renormalized thermodynamic quantities and the corresponding asymptotic limits. Finally, in Section V, we present our conclusions.
%

%%%%%%%%%%%%%%%%%%%%%%%%%%%%%%%%%%%%%%%%%%%%%%%%%%
\section{Einstein Universe and generalized zeta function method {\bf to calculate} thermal corrections}
%%%%%%%%%%%%%%%%%%%%%%%%%%%%%%%%%%%%%%%%%%%%%%%%%%
%
In this section we present the spacetime we intend to work with, i.e., the Einstein universe with a spherical boundary, as well as go through some elements of the generalized zeta function method to obtain temperature corrections to the vacuum energy at zero temperature. This Einstein universe is different from the one considered in Refs. \cite{Bezerra:2011zz, Bezerra:2011nc} in the sense that now it is viewed as a geometry inside a spherical boundary as suggested by the authors in Ref. \cite{Bayin:1993yq} (see also \cite{Ozcan:2001cr}). The topology of the spacetime is $S^3\times R^1$, but with $S^3$ equatorially bounded by $S^2$ \cite{Kennedy:1980kc}. The spherical boundaries considered in Refs. \cite{Bayin:1993yq, Kennedy:1980kc} are the ones characterized by Dirichlet and Neumann boundary conditions satisfied by a conformally coupled massless scalar field, as we shall see below.

%Another terminology given by the authors in Ref.[] is that the geometry with a spherical boundary is actually a half Einstein universe.

%%%%%%%%%%%%%%%%%%%%%%%%%%%%%%%%%
\subsection{Einstein universe with a spherical boundary}
%%%%%%%%%%%%%%%%%%%%%%%%%%%%%%%%%
The line element describing the Einstein universe inside a spherical boundary, in accordance with the authors in Refs. \cite{Bayin:1993yq, Ozcan:2001cr}, is given by
\begin{eqnarray}
ds^2 = c^2dt^2 - a^2\left[d\chi^2 + \sin^2\chi\left(d\theta^2 + \sin^2\theta d\varphi^2\right)\right],
\label{LE}
\end{eqnarray}
where $-\infty< t<\infty$ and $0\leq\chi\leq\frac{\pi}{2}$, $0\leq\theta\leq\pi$, $0\leq\varphi\leq 2\pi$ are dimensionless coordinates on a three-space of constant curvature +1. Note that the coordinate $\chi$ sweeps only half the range of the standard Einstein universe. The parameter $a$ is the constant scale factor characterizing the static closed metric and $c$, of course, is the speed of light. The finite spatial volume associated with the model considered here can be calculated as
\begin{eqnarray}
V &=& \int\sqrt{|g^{(3)}|}d\chi d\theta d\varphi \nonumber\\
&=&a^3\int_{0}^{\frac{\pi}{2}}\sin^2\chi d\chi\int_{0}^{\pi}\sin\theta d\theta\int_{0}^{2\pi}d\varphi=\pi^2 a^3,
\label{3Vol}
\end{eqnarray}
where $g^{(3)}$ is the determinant of the spatial part of the metric related to the line element \eqref{LE}. Note that, in the absence of the spherical boundary, the spatial volume of the Einstein universe is given by $2\pi a^3$ \cite{Bezerra:2011zz, Bezerra:2011nc, Bezerra:2014pza, Mota:2015ppk}, that is, twice the one in Eq. \eqref{3Vol}, as it should be.

We wish to consider a real massless quantum scalar field, $\phi$, conformally coupled to the geometry codified in the metric present in the line element \eqref{LE}. The massless scalar field, hence, obeys the equation
\begin{eqnarray}
\Box\phi + \xi R\phi = 0,
\label{KGmassless}
\end{eqnarray}
where $\Box$ is the d'Alembertian operator in curved spacetime \cite{MotaDispiration}, $R=6a^{-2}$ is the scalar curvature for the Einstein universe geometry and $\xi$ is the curvature coupling. As has been said, a conformally invariant theory is to be considered in the present work, in which case $\xi=\frac{1}{6}$. The solution of the Klein-Gordon equation \eqref{KGmassless} for a conformally coupled massless scalar field is given in terms of Gegenbauer functions and spherical harmonics. This has been obtained previously, for instance, in Refs. \cite{Ford:1975su, Ford:1976fn, Bayin:1993yq}. We are in fact interested in the eigenfrequencies associated with such a solution when submitted to Dirichlet and Neumann boundary conditions at $\chi=\frac{\pi}{2}$, characterizing spherical boundaries \cite{Bayin:1993yq, Kennedy:1980kc, Ozcan:2001cr} . The eigenfrequencies in this case, along with the degeneracy $g_n$, are
\begin{eqnarray}
\omega_n = \frac{nc}{a}\qquad\qquad g_n = \frac{1}{2}n(n\pm1) \qquad\qquad n=1,2,3,...,
\label{eigenD}
\end{eqnarray}
where the minus sign stands for Dirichlet and the plus sign for Neumann boundary conditions. The expression above will be useful to calculate temperature corrections to the vacuum energy at zero temperature already obtained in Refs. \cite{Bayin:1993yq, Kennedy:1980kc, Ozcan:2001cr}.

%%%%%%%%%%%%%%%%%%%%%%%%%%%%%%%%%%%%%%%%
\subsection{Zeta function method { \bf and temperature corrections}}
%%%%%%%%%%%%%%%%%%%%%%%%%%%%%%%%%%%%%%%%

Let us now briefly point out some necessary elements of the zeta function method to implement temperature corrections. In this spirit, an operator $\hat{A}_4$, in a four-dimensional Euclidean spacetime\footnote{In essence, a four-dimensional Euclidean spacetime is reached by making $t=-i\tau$ in Eqs. \eqref{LE} and \eqref{KGmassless}.}, with eigenvalues $\lambda_{\sigma}$, has an associated generalized zeta function written as \cite{Hawking1977, Elizalde1994book, Aleixo:2021cfy}

\begin{equation}
\zeta_4(s) = \sum_{\sigma}\lambda^{-s}_{\sigma},
\label{Zfun}
\end{equation}
where $\sigma$ stands for the spectrum of eigenvalues\footnote{The set of quantum modes in our case.} of $\hat{A}_4$, which may not always be discrete. Note that the expression above converges, in four dimensions, for Re$(s) > 2$ and it is regular at $s=0$. It can also be analytically extended to a function of $s$, with poles at $s=2$ and $s=1$ \cite{Elizalde1994book}. 

%{\bf For convenience, only in the present (sub)-section we adopt natural units so that $\hbar=c=1$. This does not affect the understanding of our analysis in the next sections. }

The analytic extension of the generalized zeta function defined in Eq. \eqref{Zfun} carrying temperature terms has already been known and is written as \cite{Hawking1977, Elizalde1994book, Aleixo:2021cfy}

 \begin{equation} \label{zetaMod}
        \zeta_4(s) = \frac{\beta}{\sqrt{4\pi}\Gamma(s)}\left\{\Gamma(s - 1/2)\zeta_3(s-1/2) + 2\sum_{n=1}^\infty \int^\infty_0 \eta^{s-\frac{3}{2}}e^{-\frac{(n\beta)^2}{4\eta}}\text{Tr}\left[e^{-\eta\hat{A}_3}\right]d\eta\right\},
    \end{equation}
   where $\hat{A}_3$ is a differential operator defined as the spatial part of $\hat{A}_4$, $\zeta_3(s-1/2) $ is the generalized zeta function written in terms of the eigenvalues of $\hat{A}_3$ and $\beta=\frac{1}{k_B T}$, with $k_B$ being the Boltzmann constant and $T$ the temperature.
   
The connection of a physical quantum system with thermodynamics is realized by means of the partition function, which can be shown to be given by \cite{Hawking1977, Elizalde1994book, Aleixo:2021cfy}%
  \begin{eqnarray}
        Z = \left[\text{det}\left(\frac{4}{\pi\mu^2}\hat{A}_4\right)\right]^{-\frac{1}{2}},                      
        \label{ZP}
    \end{eqnarray}
where $\mu$ is a constant with dimension of mass and the four-dimensional Euclidean operator is identified in our case as the d'Alembertian $\hat{A}_4=\Box_{\text{E}}$. Hence, in natural units, the latter has dimension of mass squared so that Eq. \eqref{ZP} is dimensionless, as it should be. Furthermore, the partition function above can also be written in terms of the zeta function \eqref{zetaMod} and its derivative with respect to $s$, both calculated at $s=0$, as follows
  \begin{eqnarray}
        \ln Z = \frac{1}{2}\zeta_4'(0) + \frac{1}{2}\ln\left(\frac{\pi\mu^2}{4}\right)\zeta_4(0),
                      \label{Zlog}
    \end{eqnarray}
where we have made use of the identity $e^{-\zeta_4'(0)}=\text{det}(\hat{A}_4)$ (see \cite{Aleixo:2021cfy} for more details). Therefore, only the first term in the r.h.s of the above expression contributes to the free energy \eqref{freeEnergy} since $\zeta_4(0)\rightarrow 0$, from Eq. \eqref{zetaMod}. As the physical result should not depend on the parameter $\mu$, this is expected.  

The free energy, as usual, is found to be \cite{Hawking1977, Elizalde1994book, Aleixo:2021cfy}
  \begin{eqnarray} \label{freeEnergy}
        F &=& - \frac{\ln Z}{\beta}\nonumber\\
        &=& \frac{1}{2}\zeta_3(-1/2) - \frac{1}{\sqrt{4\pi}}\sum_{n=1}^\infty \int^\infty_0 \eta^{-\frac{3}{2}}e^{-\frac{(n\beta)^2}{4\eta}}\text{Tr}\left[e^{-\eta\hat{A}_3}\right]d\eta.
    \end{eqnarray}
Furthermore, if one assumes that $\epsilon^2_j$ are the eigenvalues of the three-dimensional spatial operator, $\hat{A}_3$, then, the trace above is given by
      \begin{eqnarray} \label{trace}
       \text{Tr}\left[e^{-\eta\hat{A}_3}\right]=\sum_je^{-\epsilon^2_j\eta},
    \end{eqnarray}
where $j$ stand for the quantum modes associated with the operator $\hat{A}_3$. Hence, by performing the integration in $\eta$ present in Eq. (\ref{freeEnergy}), and taking the sum in $n$ afterwards, we obtain
  \begin{eqnarray} \label{freeEnergy2}
        F^{\text{ren}} &=& E_0^{\text{ren}} + F_T\nonumber\\
        &=&\frac{1}{2}\zeta_3(-1/2) +\frac{1}{\beta}\sum_{j}\ln\left(1-e^{-\epsilon_j\beta}\right),
    \end{eqnarray}
where the superscript `ren' stands for renormalized.

Note that, in our case, $\epsilon_j=\hbar\omega_j$, where $\omega_j$ stands for the eigenfrequencies, \eqref{eigenD}, associated with the conformally coupled massless scalar field subjected to spherical boundary conditions and $\hbar$ is the Planck's constant. The operator $\hat{A}_3$ is the spatial part of Eq. \eqref{KGmassless} plus the curvature coupling. Note also that the zeta function $\zeta_3(s)$, through the parameter $s$, already regularizes and consequently provides a renormalized vacuum energy at zero temperature given by the first term on the r.h.s. of Eq. \eqref{freeEnergy2}. The second term on the r.h.s., though, accounts for temperature corrections.

At high temperatures the correction $F_T$, obtained in a given restricted volume $V$, may contain terms of quantum nature with the following structure:
\begin{eqnarray}
\alpha_0\frac{(k_BT)^4}{(\hbar c)^3},\qquad\qquad\alpha_1\frac{(k_BT)^3}{(\hbar c)^2}, \qquad\qquad\alpha_2\frac{(k_BT)^2}{\hbar c},
\label{QNterms}
\end{eqnarray}
where $\alpha_0, \alpha_1, \alpha_2$ have, respectively, dimensions of cm$^3$, cm$^2$ and cm and are given in terms of the heat kernel coefficients \cite{Bezerra:2011zz, Bezerra:2011nc, Geyer:2008wb, Dowker:1978md}. It has been argued that these terms should be subtracted from $F_T$ in order to provide the dominant and correct classical limit at high temperatures, which is independent of Planck's constant. This is in accordance with investigations conducted for material bodies, within the framework of Lifshitz theory \cite{Lifshitz:1956zz, Bezerra:2011zz, Bezerra:2011nc, Geyer:2008wb}. Thereby, $F_T$ must go through a finite renormalization process. This subtraction is also motivated by the fact that the coefficients $\alpha_0, \alpha_1, \alpha_2$ have the same geometric structure as in the corresponding infinite terms to be subtracted from the vacuum energy at zero temperature, as shown for instance in Ref. \cite{Geyer:2008wb}, in the case of retangular boxes. In the case of the Einstein universe considering the conformally coupled massless scalar field investigated in Ref. \cite{Bezerra:2011zz}, it has been shown that the only contribution is the one for which $\alpha_0\neq 0$, that is,
\begin{eqnarray}
\alpha_0 = -\frac{\pi^2}{90}V,\qquad\qquad\alpha_1=0,\qquad\qquad\alpha_2=0.
\label{HKcoef}
\end{eqnarray}
However, as we shall see in the next section, in the case of a conformally coupled massless scalar field in the Einstein universe with spherical boundaries, there also exists an additional contribution for which $\alpha_1\neq 0$, besides the one in Eq. \eqref{HKcoef}, associated with the black-body radiation. In this case
\begin{eqnarray}
\alpha_0 = -\frac{\pi^2}{90}V,\qquad\qquad\alpha_1= \mp\frac{\zeta(3)a^2}{2},\qquad\qquad\alpha_2=0,
\label{HKcoef2}
\end{eqnarray}
where we should emphasize that the up sign always stands for Neumann boundary condition. Therefore, in our case, it should be subtracted from $F_T$ not only the black-body radiation contribution but also the contribution proportional to $(k_BT)^3$, with coefficient $\alpha_1\neq 0$ given above.

The free energy in Eq. \eqref{freeEnergy2} allows us to calculate other thermodynamics quantities such as the internal energy given by
\begin{equation}
U = -T^2\frac{\partial}{\partial T}\left(\frac{F}{T}\right),
\label{internalE}
\end{equation}
as well as the entropy
\begin{equation}
S = -\frac{\partial F}{\partial T}.
\label{entropy}
\end{equation}
Therefore, along with the free energy \eqref{freeEnergy2}, the last two thermodynamics quantities complement our analysis. In the next section we provide analytical expressions, for each one of them, suitable to investigate the limits of high and low temperatures.
%\begin{equation}
%P = -\frac{\partial F}{\partial V}
%\label{pressure}
%\end{equation}

%%%%%%%%%%%%%%%%%%%%%%%%%%%%%%%%%%%%%%%
\section{Finite renormalization terms and renormalized Casimir free energy }
%%%%%%%%%%%%%%%%%%%%%%%%%%%%%%%%%%%%%%%
%
As it has been remarked in the previous section, the first term on the r.h.s of the free energy \eqref{freeEnergy2} provides the renormalized vacuum energy at zero temperature. Thus, by making use of the eigenfrequencies and their degeneracy in Eq. \eqref{eigenD} we are able to show that\footnote{We call attention to the fact that $\zeta_3(s)$ is defined by \eqref{Zfun}, with $\lambda_j = \omega_j^2$ and $j=n$.}
\begin{eqnarray}\label{eingen}
    E^{\text{ren}}_0 &=& \frac{\hbar}{2}\zeta_3(-1/2)\nonumber\\
        &=& \frac{\hbar c}{480a}.
\end{eqnarray}
This result has already been obtained previously by the authors in Refs. \cite{Bayin:1993yq} by making use of the cutoff method. Here we have shown that the generalized zeta function method also provides the same result, as it should be. The result is the same for Dirichlet and Neumann boundary conditions. This is because the term $n$ in the degeneracy gives a contribution that is $\zeta(-2)=0$. Note that the vacuum energy given above is half of the one for the Einstein universe without spherical boundary considered in Refs. \cite{Bezerra:2011zz, Bezerra:2011nc, Bezerra:2014pza, Mota:2015ppk}.

As to temperature corrections, we need to investigate the second term on the r.h.s of the free energy \eqref{freeEnergy2}, taking into account the eigenfrequencies in \eqref{eigenD}, that is,
\begin{eqnarray}\label{FreeNonReg2}
    F_T=\frac{1}{\beta}\sum_{n=1}^{\infty}g_n\ln\left(1-e^{-\hbar\omega_n\beta}\right).
\end{eqnarray}

In order to perform the sum in $n$ above we can make use of the Abel-Plana formula \cite{Saharian:2000xx}
\begin{equation}\label{Abel-Plana}
    \sum_{n=1}^{\infty}\Phi(n)-\int_{0}^{\infty}\Phi(t)dt=-\frac{1}{2}\Phi(0)+i\int_{0}^{\infty}\frac{\Phi(i t)-\Phi(-i t)}{\exp{(2\pi t)-1}}dt.
\end{equation}
For our discussion we use Eq. \eqref{FreeNonReg2} to make the following identification:
\begin{equation}
\Phi(n) = \frac{k_BT}{2}\sum_{n=1}^{\infty}(n^2\pm n)\ln\left(1-e^{-\frac{\hbar cn}{ak_BT}}\right).
\label{funPHI}
\end{equation}
As a consequence, $\Phi(0)=0$, and the integral in the l.h.s of the Abel-Plana formula \eqref{Abel-Plana} provides
\begin{eqnarray}\label{int1}
    \int_0^{\infty}\Phi(t) = -\frac{\pi^2}{90}V\frac{(k_BT)^4}{(\hbar c)^3} \mp \frac{\zeta(3)a^2}{2}\frac{(k_BT)^3}{(\hbar c)^2},
\end{eqnarray}
where the spatial volume $V$ is given by Eq. \eqref{3Vol}. We can see that the first term on the r.h.s is the black-body radiation contribution, also present for the Einstein universe case without boundary as analyzed in Ref. \cite{Bezerra:2011zz}. Here, as a consequence of the spherical boundaries, a second contribution comes about, given by the second term in the r.h.s. of Eq. \eqref{int1}. As discussed in Ref. \cite{Bezerra:2011zz}, and already mentioned in the previous section, these two contributions must be subtracted from \eqref{FreeNonReg2}, allowing us to adopt a finite renormalization process that provides the correct classical limit at high temperatures.

Let us then define the renormalized temperature correction term by subtracting the contributions in Eq. \eqref{int1} from $F_T$ in Eq. \eqref{FreeNonReg2}, i.e.,
\begin{eqnarray}\label{RFEnergy}
F^{\text{ren}}_{T} &=&  \sum_{n=1}^{\infty}\Phi(n)-\int_{0}^{\infty}\Phi(t)dt\nonumber\\
&=& i\int_{0}^{\infty}\frac{\Phi(i t)-\Phi(-i t)}{\exp{(2\pi t)-1}}dt.
\end{eqnarray}
Now, by using the following series expansion for the logarithmic function,
\begin{eqnarray}\label{series}
    \ln(1-z) = - \sum_{n=1}^{\infty}\frac{z^n}{n},
\end{eqnarray}
we are able to obtain, by considering Eq. \eqref{funPHI}, part of the integrand present in the r.h.s of \eqref{RFEnergy}, in the form
\begin{eqnarray}
\Phi(it) - \Phi(-it) = &-&ik_BT t^2\sum_{n=1}^{\infty}\frac{1}{n}\sin\left(\frac{\hbar cnt}{ak_BT}\right)\nonumber\\
&\mp& ik_BT t\sum_{n=1}^{\infty}\frac{1}{n}\cos\left(\frac{\hbar cnt}{ak_BT}\right).
\label{difPHI}
\end{eqnarray}
Hence, the renormalized temperature correction, \eqref{RFEnergy}, can be re-written as

\begin{eqnarray}
F^{\text{ren}}_{T} &=& k_BT\sum_{n=1}^{\infty}\frac{1}{n}\int_{0}^{\infty}\frac{t^2\sin\left(\frac{\hbar cnt}{ak_BT}\right)}{\exp{(2\pi t)-1}}dt\nonumber\\
&\pm&k_BT \sum_{n=1}^{\infty}\frac{1}{n}\int_{0}^{\infty}\frac{t\cos\left(\frac{\hbar cnt}{ak_BT}\right)}{\exp{(2\pi t)-1}}dt.
\label{renFE}
\end{eqnarray}
The result above, thus, allows us to obtain the renormalized Casimir free energy of the conformally coupled massless scalar field in the Einstein universe with spherical boundaries as given by
\begin{eqnarray}\label{renCFE}
   F^{\text{ren}}_{\text{C}} = \frac{\hbar c}{480a} + F^{\text{ren}}_{T}.
\end{eqnarray}
This is in fact the expression in Eq. \eqref{freeEnergy2} when the contributions in Eq. \eqref{int1} are subtracted from $F_T$. Note that we could straightforwardly solve the integrals in Eq. \eqref{renFE}, which would lead to an expression adequately convenient only to analyze the low-temperature limit of the renormalized Casimir free energy defined above. Instead, we wish to obtain a result that is able to provide us with appropriate expressions to consider the limits of both, high and low temperatures, something that will be made in the next section.
%\begin{eqnarray}
%F^{\text{ren}}_{T} &=& \frac{\pi^2}{90}V\frac{(k_BT)^4}{(\hbar c)^3} - \frac{k_BT}{8}\sum_{n=1}^{\infty}\frac{1}{n}\coth\left(\frac{\hbar c\beta n}{2a}\right)\text{csch}^2\left(\frac{\hbar c\beta n}{2a}\right)\nonumber\\
%
%&\pm&\frac{\zeta(3)a^2}{2}\frac{(k_BT)^3}{(\hbar c)^2} \mp \frac{k_BT}{8}\sum_{n=1}^{\infty}\frac{1}{n}\text{csch}^2\left(\frac{\hbar c\beta n}{2a}\right)
%\label{renFE2}
%\end{eqnarray}
%%%%%%%%%%%%%%%%%%%%%%%%%%%%%%%%%%%%%%%%
\section{Renormalized thermodynamic quantities and asymptotic limits}
%%%%%%%%%%%%%%%%%%%%%%%%%%%%%%%%%%%%%%%%
In this section we are interested in deriving analytic expressions for the renormalized Casimir free energy that will allow us more conveniently to investigate the limits of low and high temperatures. In order to do that let us firstly consider the identity
\begin{eqnarray}\label{iden}
  \frac{1}{e^{2\pi t}-1}=\sum_{\ell=1}^{\infty}e^{-2\pi\ell t}.
\end{eqnarray}
This makes possible to write Eq. \eqref{renFE} alternatively as

\begin{eqnarray}
F^{\text{ren}}_{T} &=& k_BT\sum_{n=1}^{\infty}\frac{1}{n}\sum_{\ell=1}^{\infty}\int_{0}^{\infty}t^2\sin\left(\frac{\hbar cnt}{ak_BT}\right)e^{-2\pi \ell t}dt\nonumber\\
&\pm&k_BT \sum_{n=1}^{\infty}\frac{1}{n}\sum_{\ell=1}^{\infty}\int_{0}^{\infty}t\cos\left(\frac{\hbar cnt}{ak_BT}\right)e^{-2\pi \ell t}dt.
\label{renFE2}
\end{eqnarray}
Consequently, the integrals above can be performed and the result is given in terms of a double sum by
\begin{eqnarray}\label{completeexpression}
F^{\text{ren}}_T&=&\frac{2\hbar c}{a} \sum_{n=1}^{\infty}\sum_{\ell=1}^{\infty}\frac{3(2\pi\ell)^2 - \left(\frac{\hbar c n}{ak_BT}\right)^2}{\left[(2\pi\ell)^2 + \left(\frac{\hbar c n}{ak_BT}\right)^2\right]^3} \pm k_BT\sum_{n=1}^{\infty}\frac{1}{n}\sum_{\ell=1}^{\infty}\frac{(2\pi\ell)^2 - \left(\frac{\hbar c n}{ak_BT}\right)^2}{\left[(2\pi\ell)^2 + \left(\frac{\hbar c n}{ak_BT}\right)^2\right]^2}.
\end{eqnarray}
This is an exact and convergent result. As we shall see below, by performing first the sum in $\ell$ we obtain a convenient expression to investigate the low-temperature limit of \eqref{completeexpression}. Conversely, by performing first the sum in $n$ we obtain a convenient expression to investigate the high-temperature limit. Note that the first term on the r.h.s has already been analyzed\footnote{This term comes from the $\frac{n^2}{2}$ part of the degeneracy in Eq. \eqref{eigenD}. Apart from the factor 1/2, $n^2$ is exactly the degeneracy of the vacuum energy in the Einstein universe considered in Ref. \cite{Bezerra:2011zz}.} in Ref. \cite{Bezerra:2011zz} and we do not intend to revisit this again. We will only quote the results when necessary. We are in fact interested in investigating the effects of the second term in the r.h.s. of Eq. \eqref{completeexpression}, which from now on we denote as
\begin{equation}\label{FreeNonReg4}
    \Delta F_T^{\text{ren}}=\pm k_BT\sum_{n=1}^{\infty}\frac{1}{n}\sum_{\ell=1}^{\infty}\frac{(2\pi\ell)^2 - \left(\frac{\hbar c n}{ak_BT}\right)^2}{\left[(2\pi\ell)^2 + \left(\frac{\hbar c n}{ak_BT}\right)^2\right]^2}.
\end{equation}
This nonzero temperature contribution arises from the $\frac{n}{2}$ part of the degeneracy in Eq. \eqref{eigenD}, which as we have seen does not contribute to the vacuum energy at zero temperature but it is somewhat relevant for temperature corrections.

Let us firstly perform the summation in $\ell$ now. For this purpose we may re-write Eq. \eqref{FreeNonReg4} as
\begin{equation}\label{FreeNonReg4low1}
    \Delta F_T^{\text{ren}}=\pm \frac{k_BT}{(2\pi)^2}\sum_{n=1}^{\infty}\frac{1}{n}\sum_{\ell=1}^{\infty}\frac{\ell^2 - \frac{n^2}{(2\pi\nu)^2}}{\left[\ell^2 + \frac{n^2}{(2\pi\nu)^2}\right]^2},
\end{equation}
where we have defined the dimensionless parameter $\nu\equiv\frac{ak_BT}{\hbar c}$. Consequently, the summation in $\ell$ can be performed by using the following formula:
\begin{equation}\label{formula1}
 \sum_{\ell=1}^{\infty}\frac{\ell^2 - q^2}{\left(\ell^2 + q^2\right)^2} = \frac{1}{2q^2}\left[1 - q^2\pi^2\text{csch}^2(\pi q)\right],
\end{equation}
where in our case $q=\frac{n}{2\pi\nu}$. Thus, using the result given by Eq. \eqref{formula1} in Eq. \eqref{FreeNonReg4low1}, we find
\begin{eqnarray}\label{FreeNonReg4L}
    \Delta F_T^{\text{ren}}=\pm\frac{\zeta(3)a^2}{2}\frac{(k_BT)^3}{(\hbar c)^2} \mp \frac{k_BT}{8}\sum_{n=1}^{\infty}\frac{1}{n}\text{csch}^2\left(\frac{\hbar c n}{2ak_BT}\right),
\end{eqnarray}
where $\zeta(s)$ is the Riemann zeta function, calculated at $s=3$ \cite{Elizalde1994book}. The corresponding renormalized internal energy associated with the temperature contribution in Eq. \eqref{FreeNonReg4L} is obtained by making use of Eq. \eqref{internalE}, i.e.,
\begin{eqnarray}\label{InternalErenlow}
    \Delta U^{\text{ren}}=\mp\zeta(3)a^2\frac{(k_BT)^3}{(\hbar c)^2} \pm \frac{\hbar c}{8a}\sum_{n=1}^{\infty}\coth\left(\frac{\hbar c n}{2ak_BT}\right)\text{csch}^2\left(\frac{\hbar c n}{2ak_BT}\right).
\end{eqnarray}
In addition, from Eq. \eqref{entropy}, we can also calculate the corresponding renormalized entropy associated with \eqref{FreeNonReg4L}. This is given by
\begin{eqnarray}\label{entropylow}
    \Delta S^{\text{ren}}=\mp\frac{3\zeta(3)k_B}{2}a^2\frac{(k_BT)^2}{(\hbar c)^2} \pm \frac{k_B}{8}\sum_{n=1}^{\infty}\text{csch}^2\left(\frac{\hbar c n}{2ak_BT}\right)\left[\frac{1}{n} + \frac{\hbar c}{ak_BT}\coth\left(\frac{\hbar c n}{2ak_BT}\right)\right].
\end{eqnarray}

Let us consider again Eq. \eqref{FreeNonReg4} now written in the form
\begin{equation}\label{FreeNonReg4high}
    \Delta F_T^{\text{ren}}=\pm k_BT\nu^2\sum_{\ell=1}^{\infty}\sum_{n=1}^{\infty}\frac{1}{n}\frac{(2\pi\nu\ell)^2 - n^2}{\left[(2\pi\nu\ell)^2 + n^2\right]^2}.
\end{equation}
Thereby, the summation in $n$ is performed by using the formula
\begin{equation}\label{formula2}
 \sum_{n=1}^{\infty}\frac{1}{n}\frac{p^2 - n^2}{\left(p^2 + n^2\right)^2} = \frac{1}{2p^2}\left\{2\gamma_{\text{E}} + \psi\left(1-ip\right) + \psi\left(1+ip\right) + ip\left[\psi^{(1)}\left(1-ip\right) - \psi^{(1)}\left(1+ip\right)\right]\right\},
\end{equation}
where $\gamma_{\text{E}} \simeq 0.577216 $ is the Euler's constant and $\psi^{(k)}(z)$ are the polygamma functions, with $\psi^{0}(z)=\psi(z)$ being the digamma function \cite{abramowitz, gradshteyn2000table}. Thus, since in our case $p=2\pi\nu\ell$, Eqs. \eqref{FreeNonReg4high} and \eqref{formula2} provide
\begin{eqnarray}\label{FreeNonReg4high2}
    \Delta F_T^{\text{ren}}=\pm \frac{\gamma_{\text{E}}}{24}k_BT &\pm& \frac{k_BT}{8\pi^2}\sum_{\ell=1}^{\infty}\frac{1}{\ell^2}\left\{\psi\left(1-i2\pi\nu\ell\right) + \psi\left(1+i2\pi\nu\ell\right)\right.\nonumber\\
    &&\left. + i2\pi\nu\ell\left[\psi^{(1)}\left(1-i2\pi\nu\ell\right) - \psi^{(1)}\left(1+i2\pi\nu\ell\right)\right]\right\}.
\end{eqnarray}
We can note that there exists a classical contribution proportional to $k_BT$ in the exact expression above, something that we would expect to appear only in the high-temperature limit. Moreover, the corresponding internal energy, from Eqs. \eqref{internalE} and \eqref{FreeNonReg4high2} is found to be
\begin{eqnarray}\label{InternalErenhigh}
    \Delta U^{\text{ren}}=\mp\frac{1}{2}\frac{(k_BT)^3}{(\hbar c)^2}a^2\sum_{\ell=1}^{\infty}\left[\psi^{(2)}\left(1-i2\pi\nu\ell\right)  + \psi^{(2)}\left(1+i2\pi\nu\ell\right) \right].
\end{eqnarray}
Likewise, the associated entropy is obtained by considering Eqs. \eqref{entropy} and \eqref{FreeNonReg4high2}, which gives us the following result
\begin{eqnarray}\label{entropyhigh}
    \Delta S^{\text{ren}}=&\mp&\frac{k_B}{8\pi^2}\sum_{\ell=1}^{\infty}\frac{1}{\ell^2}\left\{2\gamma_{\text{E}} + \psi\left(1-i2\pi\nu\ell\right) + \psi\left(1+i2\pi\nu\ell\right)   \right.\nonumber\\
   & &\left.+i2\pi\nu\ell\left[\psi^{(1)}\left(1-i2\pi\nu\ell\right) - \psi^{(1)}\left(1+i2\pi\nu\ell\right)\right]\right.\nonumber\\
   & &\left.+ (2\pi\nu\ell)^2\left[\psi^{(2)}\left(1-i2\pi\nu\ell\right) + \psi^{(2)}\left(1+i2\pi\nu\ell\right)\right] \right\}.
    \end{eqnarray}
It is important to keep in mind that the expressions \eqref{FreeNonReg4L} and \eqref{FreeNonReg4high2} are equivalent, which is equally true for the corresponding expressions for the internal energy and entropy derived above. Nevertheless, the expression \eqref{FreeNonReg4L} is convenient to investigate the low-temperature limit whilst the expression \eqref{FreeNonReg4high2} to investigate the high-temperature limit. The results derived above, associated with the temperature contribution \eqref{FreeNonReg4}, are to be considered along with the results obtained in Ref. \cite{Bezerra:2011zz} for the first term on the r.h.s. of \eqref{completeexpression} to calculate the renormalized Casimir free energy \eqref{renCFE}. Due to its lengthy nature, we will not write the whole expression here. The analysis of the asymptotic limits at high and low temperatures is to be considered in the next (sub-) sections.

To end the discussion of the thermodynamics quantities let us mention another one, namely, the pressure, calculated as minus the derivative of the renormalized Casimir free energy with respect to the volume of the spacetime. In analogy with what has bee done in Refs. \cite{Bezerra:2011zz, Bezerra:2011nc} it is straightforward to show in the case we are considering that an equation of state is also satisfied, i.e.,
\begin{equation}
P_{\text{C}} = \frac{\epsilon^{\text{ren}}}{3},
\end{equation}
where $\epsilon^{\text{ren}} = \frac{U_{\text{C}}^{\text{ren}}}{V}$ is the renormalized internal energy density, with $V$ given by Eq. \eqref{3Vol}.
%
%%%%%%%%%%%%%%%%%%%%%%%%%%%
\subsection{Low-temperature limit}
%%%%%%%%%%%%%%%%%%%%%%%%%%%
The first limiting case to be considered is the one of low temperature, that is, $k_BT\ll \frac{\hbar c}{a}$. In this case, the most convenient expressions to perform this analysis are given in Eqs. \eqref{FreeNonReg4L}, \eqref{InternalErenlow} and \eqref{entropylow}. Hence, the mentioned asymptotic limit for the free energy temperature contribution in Eq. \eqref{FreeNonReg4L} is written as
\begin{equation}\label{FreeNonReg4lowT}
    \Delta F_T^{\text{ren}}\simeq \pm\frac{\zeta(3)a^2}{2}\frac{(k_BT)^3}{(\hbar c)^2} \mp \frac{k_BT}{2}e^{-\frac{\hbar c }{ak_BT}}.
\end{equation}
Therefore, the low-temperature limit for the renormalized Casimir free energy takes into consideration the renormalized free energy in Eq. \eqref{completeexpression}. In the latter, the low-temperature limit for the first term in the r.h.s has been considered in Ref. \cite{Bezerra:2011zz}. Thus, the low-temperature limit for the renormalized Casimir free energy \eqref{renCFE} is written as
\begin{eqnarray}\label{FreeNonReg4lowT2}
 F^{\text{ren}}_{\text{C}} &\simeq&  \frac{\hbar c}{480a} + \frac{\pi^2}{90}V\frac{(k_BT)^4}{(\hbar c)^3} - \frac{k_BT}{2}e^{-\frac{\hbar c }{ak_BT}}\nonumber\\
 &\pm&\frac{\zeta(3)a^2}{2}\frac{(k_BT)^3}{(\hbar c)^2} \mp \frac{k_BT}{2}e^{-\frac{\hbar c }{ak_BT}}.
\end{eqnarray}
Note, however, that the low-temperature limit of the first term in the r.h.s of Eq. \eqref{completeexpression} has been considered in Ref. \cite{Bezerra:2011zz} without the 1/2 factor coming from the degeneracy $g_n$ in Eq. \eqref{eigenD}. Thus, the first line in Eq. \eqref{FreeNonReg4lowT2} is just the result from Ref. \cite{Bezerra:2011zz} multiplied by 1/2. Another important point to note is that, in the case of Dirichlet boundary condition, the exponentials above cancel out, surviving only the terms proportionals to $(k_BT)^4$ and $(k_BT)^3$ as corrections to the vacuum energy at zero temperature, in the first term in the r.h.s. of Eq. \eqref{FreeNonReg4lowT2}. Conversely, in the case of Neumann boundary condition, the exponentially suppressed terms sum up, giving the same order exponential correction as in Ref. \cite{Bezerra:2011zz}.

The discussion above follows in a similar way for the Casimir internal energy obtained from Eqs. \eqref{internalE} and \eqref{completeexpression}. Hence, by taking the low-temperature limit of Eq. \eqref{InternalErenlow} and considering the result from Ref. \cite{Bezerra:2011zz}, we obtain
\begin{eqnarray}\label{CasimirIE}
 U^{\text{ren}}_{\text{C}} &\simeq&  \frac{\hbar c}{480a} - \frac{\pi^2}{30}V\frac{(k_BT)^4}{(\hbar c)^3} + \frac{\hbar c}{2a}e^{-\frac{\hbar c }{ak_BT}}\nonumber\\
 &\mp&\frac{\zeta(3)a^2}{2}\frac{(k_BT)^3}{(\hbar c)^2} \pm \frac{\hbar c}{2a}e^{-\frac{\hbar c }{ak_BT}},
\end{eqnarray}
which is also dominated by the vacuum energy at zero temperature, as we can see.
%The second line of the above result can also be inferred directly from Eq. \eqref{FreeNonReg4lowT}, by making use of Eq. \eqref{internalE}.

Finally, to end our analysis of the low-temperature limit let us consider the Casimir entropy. This can be reached by making use of Eqs. \eqref{entropy} and \eqref{completeexpression}. Thus, by taking into consideration the low-temperature limit of Eq. \eqref{entropylow} and considering the result from Ref. \cite{Bezerra:2011zz}, we found
\begin{eqnarray}\label{CasimirE}
 S^{\text{ren}}_{\text{C}} &\simeq& - \frac{2\pi^2}{45}k_B\left(\frac{k_BT}{\hbar c}\right)^3V + \frac{\hbar c}{2aT}e^{-\frac{\hbar c }{ak_BT}}\nonumber\\
 &\mp&\frac{\zeta(3)a^2}{2}\frac{(k_BT)^3}{(\hbar c)^2} \pm \frac{\hbar c}{2aT}e^{-\frac{\hbar c }{ak_BT}}.
\end{eqnarray}
Again, for Dirichlet boundary condition the small exponential corrections cancel out, while for Neumann boundary condition they sum up. We can notice that in the limit $T\rightarrow 0$ the Casimir entropy goes to zero and the third law of thermodynamics is satisfied, according to Nernst heat theorem \cite{Landau, Bezerra:2011zz}.
%
%%%%%%%%%%%%%%%%%%%%%%%%%%%%%%%%%%%%%%%%%%%%%
\subsection{High-temperature limit}
%%%%%%%%%%%%%%%%%%%%%%%%%%%%%%%%%%%%%%%%%%%%%
We now turn to the high-temperature limit case, i.e., $k_BT\gg \frac{\hbar c}{a}$. The most convenient expressions for this investigation are the ones in Eqs. \eqref{FreeNonReg4high2}, \eqref{InternalErenhigh} and \eqref{entropyhigh}. Thereby, the asymptotic limit for high temperatures of the renormaized free energy contribution in Eq. \eqref{FreeNonReg4high2} is given by the dominant terms
\begin{eqnarray}\label{FreeNonReg4khighT}
    \Delta F_T^{\text{ren}}\simeq \mp \frac{k_BT}{24}\left(1 - 12\ln A\right) \pm \frac{k_BT}{24}\ln\left(\frac{k_BTa}{\hbar c}\right) \pm\frac{1}{5760}\left(\frac{\hbar c}{a}\right)^2\frac{1}{k_BT},
\end{eqnarray}
where $A\simeq 1.28$ is the Glaisher's constant and the largest (small) correction to the linear and logarithmic terms has been written. Thus, the renormalized Casimir free energy is obtained, from Eq. \eqref{renCFE},  by making use of Eq. \eqref{completeexpression}, the result of Ref. \cite{Bezerra:2011zz} for high temperatures and the expression above. This gives
\begin{eqnarray}\label{FreeNonReg4khighTC}
     F_{\text{C}}^{\text{ren}}&\simeq& \frac{k_BT}{8\pi^2}\zeta(3) + 2\pi^2\left(\frac{ak_BT}{\hbar c}\right)^2\frac{\hbar c}{a}e^{-4\pi^2\frac{ak_BT}{\hbar c}}\nonumber\\
     &\pm& \frac{k_BT}{24}\left(12\ln A - 1\right) \pm \frac{k_BT}{24}\ln\left(\frac{k_BTa}{\hbar c}\right) \pm\frac{1}{5760}\left(\frac{\hbar c}{a}\right)^2\frac{1}{k_BT}.
\end{eqnarray}
Note that the first term in the r.h.s of Eq. \eqref{renCFE} has been canceled with the same contribution, but with opposite sign, coming from the high-temperature limit of the first term in the r.h.s. of \eqref{completeexpression}  \cite{Bezerra:2011zz}. We can also notice that, compared with the results in Ref. \cite{Bezerra:2011zz} for the Einstein universe without boundaries, the asymptotic expression above brings one more classical term, as well as a logarithmic term similar to the electromagnetic case analyzed in Ref. \cite{Bezerra:2011nc}. Moreover, the largest correction to the linear and logarithmic terms is the one proportional to $\frac{1}{k_BT}$, not the exponential as it is the case in Ref. \cite{Bezerra:2011zz}. It is worth pointing out that, near the singularity ($a\to 0$), the logarithmic term dominates over the other ones, with both thermal and quantum features of the primordial microscopic Universe competing with each other. In this case, the quantum features dominate, as expected.

The Casimir internal energy at high temperatures can be obtained by considering the result from Ref. \cite{Bezerra:2011zz} and by using Eqs. \eqref{internalE}, \eqref{completeexpression} and \eqref{InternalErenhigh}. The complete expression is given by
\begin{eqnarray}\label{CasimirIEhigh}
 U^{\text{ren}}_{\text{C}} &\simeq&  8\pi^4\left(\frac{k_BTa}{\hbar c}\right)^4\frac{\hbar c}{a}e^{-4\pi^2\frac{ak_BT}{\hbar c}}\nonumber\\
 &\mp&\frac{k_BT}{24} \pm \frac{1}{2880}\left(\frac{\hbar c}{a}\right)^2\frac{1}{k_BT}.
\end{eqnarray}
Thereby, the expression above brings a classical contribution and a correction that is proportional to $\frac{1}{k_BT}$, which is bigger than the exponential correction in the first line. Therefore, the Casimir internal energy is completely dominated by the second line of the expression \eqref{CasimirIEhigh}. In the Einstein universe analyzed in Ref. \cite{Bezerra:2011zz}, the Casimir internal energy is exponentially suppressed.

To end the analysis of the high-temperature limit, we consider the Casimir entropy. The latter can be obtained by using the result of Ref. \cite{Bezerra:2011zz}, and Eqs. \eqref{entropy}, \eqref{completeexpression} and \eqref{entropyhigh}. The Casimir entropy at high temperatures is, then, written as
\begin{eqnarray}\label{entropyhigh2}
     S_{\text{C}}^{\text{ren}}&\simeq& - \frac{k_B}{8\pi^2}\zeta(3) + 8\pi^4k_B\left(\frac{ak_BT}{\hbar c}\right)^2e^{-4\pi^2\frac{ak_BT}{\hbar c}}\nonumber\\
     &\mp& \frac{k_B}{2}\ln A \mp \frac{k_B}{24}\ln\left(\frac{k_BTa}{\hbar c}\right) \mp\frac{k_B}{5760}\left(\frac{\hbar c}{ak_BT}\right)^2,
\end{eqnarray}
where we can see two constant classical terms, We can also see that the Casimir entropy is dominated by a logarithmic term of quantum origin, differently from Ref. \cite{Bezerra:2011zz} where the entropy is dominated by a constant classical term at high temperatures.

%%%%%%%%%%%%%%%%%%%%%%%%%
\section{Conclusions}
%%%%%%%%%%%%%%%%%%%%%%%%%
%
In this work we have investigated temperature corrections to the vacuum energy at zero temperature of a conformally coupled massless scalar field in the Einstein universe with a spherical boundary.  The field is subjected to obey both Dirichlet and Neumann boundary conditions. Hence, we have generalized the results found in Refs. \cite{Bayin:1993yq, Kennedy:1980kc} where the authors considered only the vacuum energy at zero temperature by adopting the cutoff method. We have then calculated the renormaized Casimir free energy, internal energy and entropy by adopting the generalized zeta function technique, as well as Abel-Plana formula.

We have also studied the limits of low and high temperatures for all thermodynamics quantities considered. In this context, we have shown that, at low temperatures, both the Casimir free energy and internal energy are dominated by the vacuum energy at zero temperature, and the entropy obeys the third law of thermodynamics in accordance with the Nernst heat theorem. Furthermore, at high temperatures we have also shown that the Casimir free energy exhibit both classical term contributions and logarithmic contributions with respect to the temperature. In this regard, the internal energy presents also a classical contribution and the entropy presents a constant classical contribution plus a logarithmic term with respect to the temperature. This is in contrast with Ref. \cite{Bezerra:2011zz} where the authors have shown that, at high temperatures, the Casimir free energy presents only a classical term, the internal energy goes to zero and the entropy is dominated by a constant classical contribution. 

As a future perspective, we intend to extend this analysis to a $D$-dimensional Einstein universe, with a hyper spherical boundary.

%
%%%%%%%%%%%%%%%%%%%%%%%%%%%%
%
%%%%%%%%%%%%%%%%%%%%%%%%%%%%%%%%%%%%%%%%%%%%%%%%%%%%%%%

%%%%%%%%%%%%%%%%%%%%%%%%%%%%%%%%%%%%%%%%%%%%%%%%%%%%%%%
%Conselho Nacional de Desenvolvimento Cient\'{i}fico e Tecnol\'{o}ogico
%

{\acknowledgments}
 The authors are partially supported by the National Council for Scientific and Technological Development (CNPq) under grants N$\textsuperscript{\underline{o}}$ 311031/2020-0 (H.F.S.M.),  308268/2021-6 (C.R.M.) and 307211/2020-7 (V.B.B.).

%\bibliography{ZetaM}
%\bibliographystyle{JHEP}
\end{document}